%% file: RADpyC-conf-proc.tex
\documentclass{rmf-d}
\usepackage{nopageno,rmfbib,multicol,times,epsf,amsmath,amssymb,cite}
\usepackage[latin1]{inputenc}
\usepackage[]{caption2}
\usepackage{graphicx}

\usepackage{siunitx}
\usepackage{float}
\usepackage{booktabs}
\usepackage{tikz}
\usetikzlibrary{arrows, arrows.meta, 3d, calc}
\usepackage{pgfplots}
\pgfplotsset{compat=1.17}
\def\rmfcornisa{Research\hfill\rmf\ {\bf ??} (*?*) ???--???
\hfill MES? A\~NO?}

\def\rmfcintilla{{\it Rev.\ Mex.\ Fis.\/} {\bf ??} (*?*) (????) ???--???}
\clearpage \rmfcaptionstyle 
\input{defs}

\begin{document}
	
\title{The First Measurement of the Muon Anomalous Magnetic Moment from the Fermilab Muon $g-2$ Collaboration
\vspace{-6pt}}
\author{A. Tewsley-Booth \emph{on behalf of the Muon $g-2$ Collaboration \thanks{https://muon-g-2.fnal.gov/collaboration.html}}}
\address{University of Kentucky, Lexington, KY, USA}
\maketitle
\recibido{day month year}{day month year
\vspace{-12pt}}

\begin{abstract}
\vspace{1em}
This paper will cover the physics and methods behind Fermilab's Muon g-2 Experiment, along with the long-awaited results from Run-1. The experiment was undertaken to resolve the tension between the Standard Model and the previous measurement taken at Brookhaven National Laboratory. The measured value of the muon magnetic anomaly is $a_\mu(FNAL)=116592040(54)\times10^{-11}$. This result is in good agreement with Brookhaven's previous measurement. The new world average, $a_\mu(Exp)=116592061(41)\times10^{-11}$, shows a difference from the theoretical value of the Standard Model (SM), $a_\mu(SM)=116591810(43)\times10^{-11}$, of 4.2 standard deviations, strongly hinting at physics beyond the Standard Model. The experiment requires the simultaneous measurement of the muon precession frequency, the magnetic field, and the muons' distribution in the field. All three of these measurements will be discussed in context, along with the main systematic corrections and uncertainties.
\vspace{1em}
\end{abstract}

\keys{Muons, Magnetic moment, Accelerators and storage rings, Precision measurements
\vspace{-4pt}}
\pacs{13.40.Em, 14.60.Ef \vspace{-4pt}}

\begin{multicols}{2}

\section{Introduction}

Our collaboration reported a new measurement of the muon magnetic anomaly, $a_\mu = \frac{1}{2}(g_\mu - 2)$, based on our Run-1 data set, which was collected between March and July of 2018. The result is \[a_\mu(\mathrm{FNAL}) = 116592040(54) \times 10^{-11},\] measured to a precision of 460 ppb. A set of four companion papers covered the final result \cite{gm2prl2021}, the anomalous spin precession frequency \cite{gm2omegaa2021}, the magnetic field measurements \cite{gm2field2021}, and the beam dynamics corrections \cite{gm2bd2021}.

This result was long awaited as a promising test of the Standard Model. The previous measurement from Brookhaven hinted at a discrepancy with the Standard Model \cite{Bennett2006}. The discrepancy, if real, is sensitive to many New Physics contributions, motivating theorists and experimentalists to improve the calculations and measurements. The goal of this new experiment at the Fermi National Accelerator Laboratory is to improve on the previous results by a factor of four. The Run-1 data set has already matched the precision of the final Brookhaven result. Runs-2 through 4 have already been completed and Run-5 is underway.

These proceedings will provide an overview of the theoretical underpinnings of the experiment, give some historical context to the measurements, and broadly cover the experimental methods and analysis.\footnote{A similar talk was given by the speaker at the NuFact2021 conference.}

\section{The Theory of Magnetic Moments}

Particles have intrinsic magnetic moments oriented along the axes of their spins. This moment, along with a particle's charge, quantifies how it behaves in an electromagnetic field. Because the magnetic moment is oriented along the spin, we can write the relationship \[\mathbf{m} = \gamma \mathbf{S},\] where $\mathbf{m}$ is the particle's magnetic moment, $\gamma = \frac{gq}{2m}$ is the gyromagnetic ratio, and $\mathbf{S}$ is the spin. This relationship leads to interesting behaviors, such as spin precession about an external non-parallel magnetic field. A magnetic moment in a magnetic field experiences a torque, \[\mathbf\tau = \mathbf{m} \times \mathbf{B}.\] The torque causes the spin vector to precess about the axis defined by the magnetic field at the Larmor frequency, \[\omega = \frac{gq}{2m} B\] \cite{Jackson}. One can see from this equation how, if $q$ and $m$ are known, simultaneous measurement of the spin precession frequency $\omega$ and the magnetic field $B$ would allow the determination of the g-factor of the particle. This method is the one used by the Muon $g-2$ Collaboration.

The Dirac equation predicts that the g-factor of a lepton such as the muon is exactly two. However, a particle's bare magnetic moment differs from its dressed moment due to interactions with the vacuum, which shift the value slightly away from two. This shift, called the magnetic anomaly $a = \frac{g-2}{2}$, is caused by interactions with virtual particles in the vacuum. Therefore, a measurement of the magnetic anomaly can be used to probe properties of these interactions. The muon is a better test ground than the electron for these measurements because its higher mass leads to an enhancement of the effect of interactions with massive virtual particles by a factor of about 43,000 \cite{gm2prl2021}.

If a particle's g-factor is exactly 2, then its spin and momentum precess at the same rate in a uniform external field, preserving its helicity as it undergoes circular motion. However, if the g-factor is not exactly 2, the spin and momentum vectors precess at slightly different rates, causing the helicity to oscillate as a relative phase accumulates between the two vectors.

Current $g-2$ theory takes into account QED, electroweak, and QCD. The QED terms dominate the value, but the QCD terms dominate the uncertainty \cite{gm2whitepaper2020}. The recent theory whitepaper uses a data driven technique to calculate hadronic sector contributions. Additionally, lattice QCD calculations are becoming more comparable, as well.

\section{Brief Historical Context}

The history of the measurement of the muon magnetic anomaly begins in 1928, when Dirac published his relativistic equation for the electron that, unexpectedly, also predicted that $g_e=2$, solving a mystery of the time. In 1948, Julian Schwinger calculated the leading order correction to the lepton g-factor, $\frac{g-2}{2} = \frac{\alpha}{2\pi}$. Schwinger's calculation was motivated by the measurement of $g_e$ by Foley and Kusch, which was consistent with a non-zero $a_e$ at the part per thousand level. This result motivated further study of radiative corrections and new experiments to investigate the predictions of the theory. It was confirmed in the muon in 1960 through a series of experiments at Nevis.

The experimental program at CERN began in 1962 and, over the course of three experiments conducted there through 1975, set the stage for the fundamental design of the experiment used by both Brookhaven and Fermilab. Two key features developed by the CERN experiments were the use of a storage ring and the choice of the muon momentum that largely cancels the effect of electric fields on the muon spin precession, allowing vertical focusing to be provided by electric fields.

In 1989, Brookhaven E821 began. This experiment improved further upon the methods of the CERN collaborations, especially in the control and measurement of the magnetic field. Then, in 2006, the Brookhaven collaboration reported hints of discrepancy with the Standard Model. Their final measurement of the muon's magnetic anomaly differed from the theoretical prediction by 3.7 $\sigma$. This tantalizing result motivated the next (and current) incarnation of the experiment at Fermilab \cite{Roberts2018}.

\section{Overview of the Measurement}

The g-factor of the muon is slightly greater than two. Therefore, in a constant magnetic field, the relative angle between the momentum and spin vectors evolves over time. The rate of change of this angle, called the anomalous precession frequency $\oa$, is (nearly) proportional to the magnetic anomaly, $\amu$. \[\vec\oa = \vec\omega_s - \vec\omega_c \approx -\frac{q}{m} \amu \vec B, \label{eq:oa-simple}\] where $\omega_s$ is the spin precession frequency and $\omega_c$ is the cyclotron (momentum precession) frequency. Simultaneous measurement of $\oa$ and the magnetic field enable determination of $\amu$.

The full form of Equation (\ref{eq:oa-simple}) is 
\begin{equation}
	\begin{split}
		\vec\oa = -\frac{q}{m} \Biggl( & \amu \vec B - \amu\frac{\gamma}{\gamma+1}(\vec\beta \cdot \vec B)\vec\beta \Biggr.\\
		&- \Biggl. \left[ \amu - \frac{1}{\gamma^2 - 1} \right] \frac{\vec \beta \times \vec E}{c} \Biggr).
	\end{split}
	\label{eq:oa-complex}
\end{equation}
This equation has two additional terms compared to the simplified form in Equation (\ref{eq:oa-simple}). The first additional term is related to the angle between the muon's momentum and the magnetic field. This term goes to zero when the field and momentum are orthogonal. In practice, there is a small vertical component to the muon momentum, which leads us to treat this term as a small correction to the measured value of $\oa$.

The second additional term is related to the electric field. In the experiment, this term can be almost entirely canceled out by choice of muon momentum (and therefore $\gamma$), called the ``magic momentum'' at $\gamma = 29.3$, $p_\mu = \SI{3.09}{\giga\electronvolt/c}$. The actual distribution of muon momenta around this value leads to another correction to $\oa$.

It is convenient to rewrite Equation (\ref{eq:oa-simple}) in terms of the product of ratios of the quantities measured in this experiment and those known from other measurements and solve for $\amu$. In this form, \[\amu = \frac{\oa}{\oppt(T_r)} \frac{\mu'_p(T_r)}{\mu_e(H)} \frac{\mu_e(H)}{\mu_e} \frac{g_e}{2} \frac{m_\mu}{m_e}, \label{eq:ratios}\] $\oa$ and $\oppt$ are the quantities specifically measured in this experiment. The magnetic field, $\oppt$, is the average magnetic field experienced by the muons (denoted by the tilde) expressed in terms of the Larmor precession frequency of a shielded proton in water (denoted by the prime) at the reference temperature $T_r$. The other pertinent ratios are the ratio between the shielded proton magnetic moment and the electron magnetic moment in hydrogen, the ratio between the electron magnetic moment in hydrogen and the bare electron magnetic moment, the electron g-factor, and the ratio of the masses of the muon and electron.

The ratio measured is this experiment, $\oa/\oppt$, can be laid out schematically as \[\frac{\oa}{\oppt} = \frac{f_\text{clock} ~\omega_\text{a, meas} ~(1 + C_e + C_p + C_{ml} + C_{pa})}{f_\text{field} ~\left\langle \omega_\mathrm{p, meas} \bigotimes \rho_\mu \right\rangle ~(1 + B_{q} + B_{k})}. \label{eq:schematical}\] In this equation, $f_\mathrm{clock}$ is the blinding factor and $f_\mathrm{field}$ is the absolute calibration of the field that takes $\omega_\text{p, meas}$ to $\opp$, the equivalent Larmor frequency of the proton in water. The term $\left\langle \omega_\mathrm{p, meas} \bigotimes \rho_\mu \right\rangle$ is the average measured field weighted by the muon distribution $\rho_\mu$. The $C$ terms in the numerator are the corrections to the measured anomalous precession frequency, which are discussed in more detail in Section \ref{sec:omega_a} The terms $B_q$ and $B_k$ are the effects of fast transient magnetic fields that cannot be measured by the field systems that measure $\omega_\text{p, meas}$.

\begin{figure}[H]
	\centering
	\begin{tikzpicture}[scale=3]
		\input{ring-schematic.tikz}
	\end{tikzpicture}
	\caption{The storage ring systems: (purple) the inflector that cancels the magnetic field in the pass-through for muon injection; (green) the 24 calorimeters that measure the decay positrons; (red) the electrostatic quadrupoles (ESQs) that provide vertical focusing to the muons; (light blue) the kickers that move the muons onto their central orbit after injection; (dark blue) the trackers that measure the muon distribution.}
	\label{fig:ring-schematic}
\end{figure}
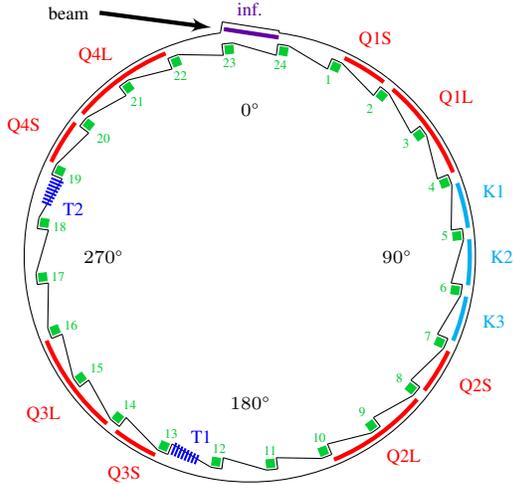

All the measurements are made in the storage ring magnet. Figure \ref{fig:ring-schematic} shows that layout of the storage ring, including the primary beam dynamics systems, the calorimeters used to measure $\oa$, and the straw trackers used to measure the muon distribution. Not shown in the figure are the field measurement systems that will be discussed in Section \ref{sec:field}

\section{Measuring $\omega_a$}
\label{sec:omega_a}

A positive muon decays into a positron and two neutrinos. The highest momentum daughter positrons occur when both neutrinos are emitted together in the opposite direction. Because the weak force that moderates the decay is parity violating, this high-momentum decay positron is preferentially emitted along the direction of the muon's spin at the time of decay. Boosted into the lab frame, this means that the highest energy daughter positrons are emitted when spin and momentum are parallel, and that there are fewer high energy positrons when the spin and momentum are anti-parallel. The positrons are detected as a function of time during a fill by an array of 24 calorimeters. Binning the hits by energy and time, we observe an oscillation in the high-energy bins that is caused by the relative phase accumulating between the muon spin and momentum. This modulation of the positron energy spectrum encodes the anomalous precession frequency, $\omega_a$, which is visible in the ``wiggle plot'' shown in Figure \ref{fig:wiggle-plot}.

Other processes also modulate the calorimeter data that need to be accounted for in order to accurately calculate $\omega_a$. Four of the most important effects are discussed here:

\begin{enumerate}
	
\item Pitch correction.

The pitch correction, $C_p$, comes from the second term in Equation (\ref{eq:oa-complex}) and is due to small vertical components of the muon momentum, leading to non-zero pitch angles. The pitch angles can be calculated by studying the vertical position distribution of the muons with the trackers because large pitch angles leads to large vertical excursions before the E-field focusing restores the muons to their equilibrium vertical position. The calculated value of the pitch correction is \SI[separate-uncertainty = true]{180(33)}{ppb}.

\begin{figure}[H]
	\centering
	\includegraphics[width=0.95\linewidth]{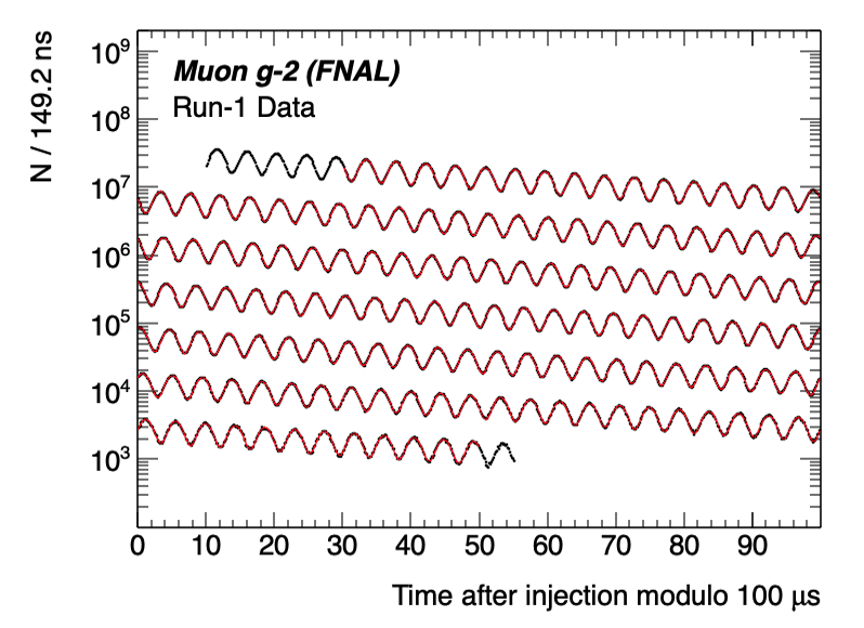}
	\caption{The modulation of high energy decay positrons encodes the anomalous precession frequency, $\oa$. This modulation rides on top of the exponential decay of the population of muons in the storage ring.}
	\label{fig:wiggle-plot}
\end{figure}

\item E-field correction.

The E-field correction, $C_e$, comes from the third term in Equation (\ref{eq:oa-complex}). Muons that are not at the magic momentum get a contribution to their anomalous precession frequency from the electric fields that provide vertical focusing. Their momentum distribution is encoded in the distribution of the equilibrium radius of the distribution. Therefore, the correction to $\oa$ from the electric field can be calculated by measuring the spread of the equilibrium radius of the beam. This correction is \SI[separate-uncertainty = true]{489(53)}{ppb}.

\item Muon loss correction.

The $\oa$ signal, which goes like $\cos(\oa t + \phi_0)$, has an initial phase $\phi_0$. If $\phi_0$ is not constant, then it can introduce an offset to the measurement of $\oa$ as $\frac{d\phi_0}{dt}$. One mechanism that can create a time-dependent phase like this is due to muon loss from the storage ring. Different momenta of muons can have different initial phases and can be lost from the ring at different rates, changing the average momentum over time. This leads to a correction that needs to be made proportional to $\frac{d\phi_0}{d\left\langle p \right\rangle} \frac{d\left\langle p \right\rangle}{dt}$. This correction is found to be \SI[separate-uncertainty = true]{-11(5)}{ppb}.

\item Phase acceptance correction.

Decay positrons do not beeline for the closest calorimeter; they can propagate in the storage ring up to about \ang{180}. This effect means that any beam motion over the course of the fill can change the relationship between a positron's decay position and its phase when detected by a calorimeter. The Run-1 data set had a significant per-fill beam motion due to a damaged resistor in one of the electric quadrupoles, making the phase acceptance effect particularly significant. It is calculated to be a correction of \SI[separate-uncertainty = true]{-158(75)}{ppb}. More recent data sets do not have the issues caused by the bad resistors, lowering this effect in future results.

\end{enumerate}

\section{Measuring $\omega_p$}
\label{sec:field}

The magnetic field experienced by the muons is measured using several different magnetometer systems. The most important of these are the calibration probe, trolley, and fixed probe systems. The primary field mapping system, the trolley, is an array of magnetometers located in the vacuum chamber. The trolley is pulled by cable around the muon storage region, measuring the field as it goes. Each of these systems uses NMR magnetometry, and together they form a calibration chain that allows us to measure the absolute field to a precision of 114 ppb (56 ppb from the calibration, measurements, analysis, and averaging; and 99 ppb from the effects of fast transient fields).

\begin{figure}[H]
	\centering
	\includegraphics[width=0.95\linewidth]{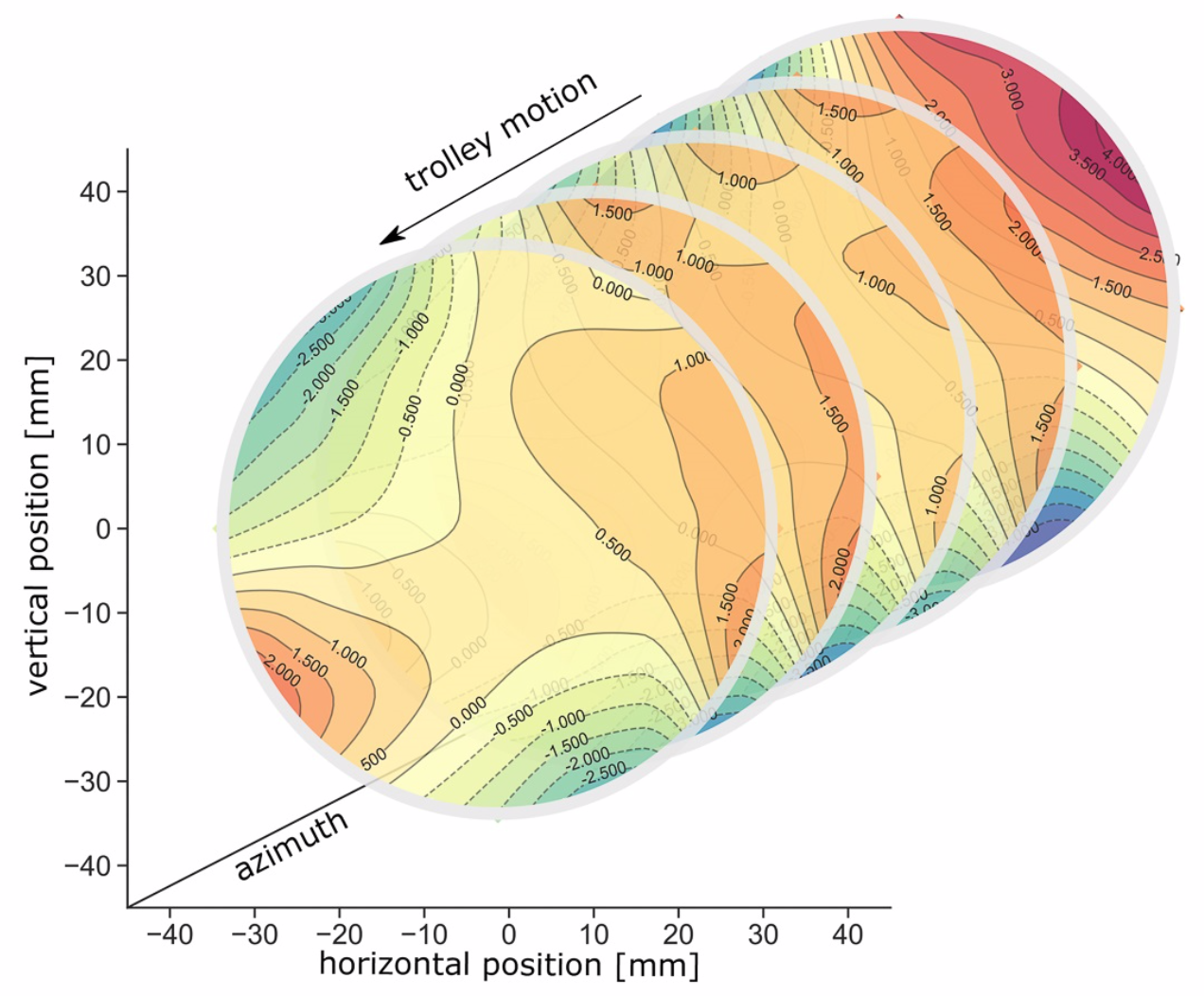}
	\caption{The trolley generates precise field maps as a function of azimuthal position as it travels around the ring. Between trolley runs, these maps are interpolated using measurements from the fixed probes that track the evolution of the field.}
	\label{fig:trolley-slices}
\end{figure}

The calibration chain begins at Argonne National Laboratory in a precision MRI magnet. There, the calibration probe is cross-checked with a He-3 NMR probe \cite{Farooq2020}. The two probes were found to be in excellent agreement. The calibration probe is then transferred to the storage ring at FNAL, where it is attached to a 3D translation stage that allows it to be inserted into the storage ring at the same positions as the trolley probes. Using a procedure called rapid swapping, the calibration from the calibration probe is transferred to each of the trolley probes. Then, as the trolley moves about the storage ring, it can transfer the calibration to the array of fixed probes.

\begin{figure}[H]
	\centering
	\includegraphics[width=0.95\linewidth]{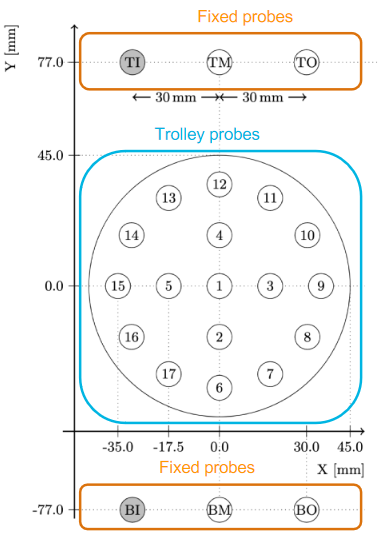}
	\caption{The relative positions of the fixed and trolley probes in an azimuthal slice of the ring. In this coordinate system, the muon's magic radius is at $(0,0)$.}
	\label{fig:probe-locations}
\end{figure}

After calibration, the primary systems used to map the magnetic field in the storage ring are the trolley and the fixed probes. The trolley is pulled through the vacuum chambers by cables and makes its measurements in the same volume that the muons fill. It has 17 NMR probes and takes measurements in about 4000 azimuthal locations around the ring, as shown in Figure \ref{fig:trolley-slices}. However, the trolley cannot be operated during muon fills because it blocks their path, so it is pulled out of the way into its garage most of the time, only mapping the field about every three days. Therefore, we say that the trolley measurements are dense in space but sparse in time. On the contrary, the fixed probes are sparse in space but dense in time; they only measure at 72 azimuthal locations and are physically located outside the vacuum chambers, but they can continue measuring the field during muon fills, providing information about how the field evolves between trolley runs. Figure \ref{fig:probe-locations} shows the relative locations of the trolley and fixed probes in an azimuthal slice of the ring. The two sources are combined in the analysis, with the fixed probe data being used to interpolate the field map between the trolley runs. For Run-1, this procedure, from measurement through analysis, accounts for 56 ppb of the total uncertainty.

The NMR magnetometers are good for measuring quasi-static fields, but transients with characteristic times faster than about a second require special systematic studies to measure. There are two primary sources of transient magnetic fields in the storage ring: the electrostatic quadrupoles (ESQs) and the faster kickers. The ESQ plates are charged to high voltage at each muon injection to provide vertical focusing on the beam. Charging the plates induces vibrations, which generate a magnetic field that perturbs the muons. For the Run-1 analysis, the effect of this transient field was found to be -17 ppb with an uncertainty of 92 ppb, making it the dominant systematic effect on the final result. Since the Run-1 publication, we have completed a more in-depth systematic study and expect this uncertainty to be reduced in future results. Furthermore, beginning in Run-5, we have changed the ESQ charging procedure to reduce the induced vibrations that cause the transient.

\begin{figure}[H]
	\centering
	\begin{tikzpicture}[scale=0.75]
		\node at (0,-0.25) {\includegraphics[width=7.5cm]{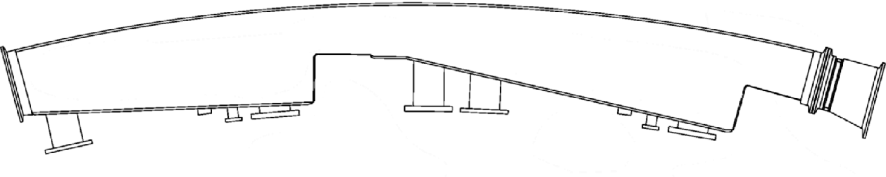}};
		\input{tracker-cartoon.tikz}
	\end{tikzpicture}
	\caption{Decay positrons travel through the straw trackers. Their paths can be traced back using knowledge of the magnetic field to their decay vertices, allowing for a calculation of the muon distribution.}
	\label{fig:tracker-cartoon}
\end{figure}

The source of the other transient is the kicker system, which uses a fast magnetic field at the beginning of each muon injection to kick the muons onto their ideal orbit. The kick induces eddy currents that perturb the field as they decay. This effect is measured using a Faraday magnetometer that can measure the field at the nanosecond level. The average effect on the muons was found to be -27 ppb, with an uncertainty of 37 ppb. This measurement has also been refined and repeated since the Run-1 publication to reduce its associated uncertainty.

\section{The Muon Distribution}

In order to calculate the average magnetic field experienced by the muons, we need to know both the magnetic field and the muon distribution in the storage ring as a function of position and time. The muon distribution is measured by the straw trackers at two azimuthal locations. Those two distributions are then used to extrapolate the distribution around the whole ring by combining the measurement with beam dynamics simulations.

\begin{figure}[H]
	\centering
	\includegraphics[width=0.95\linewidth]{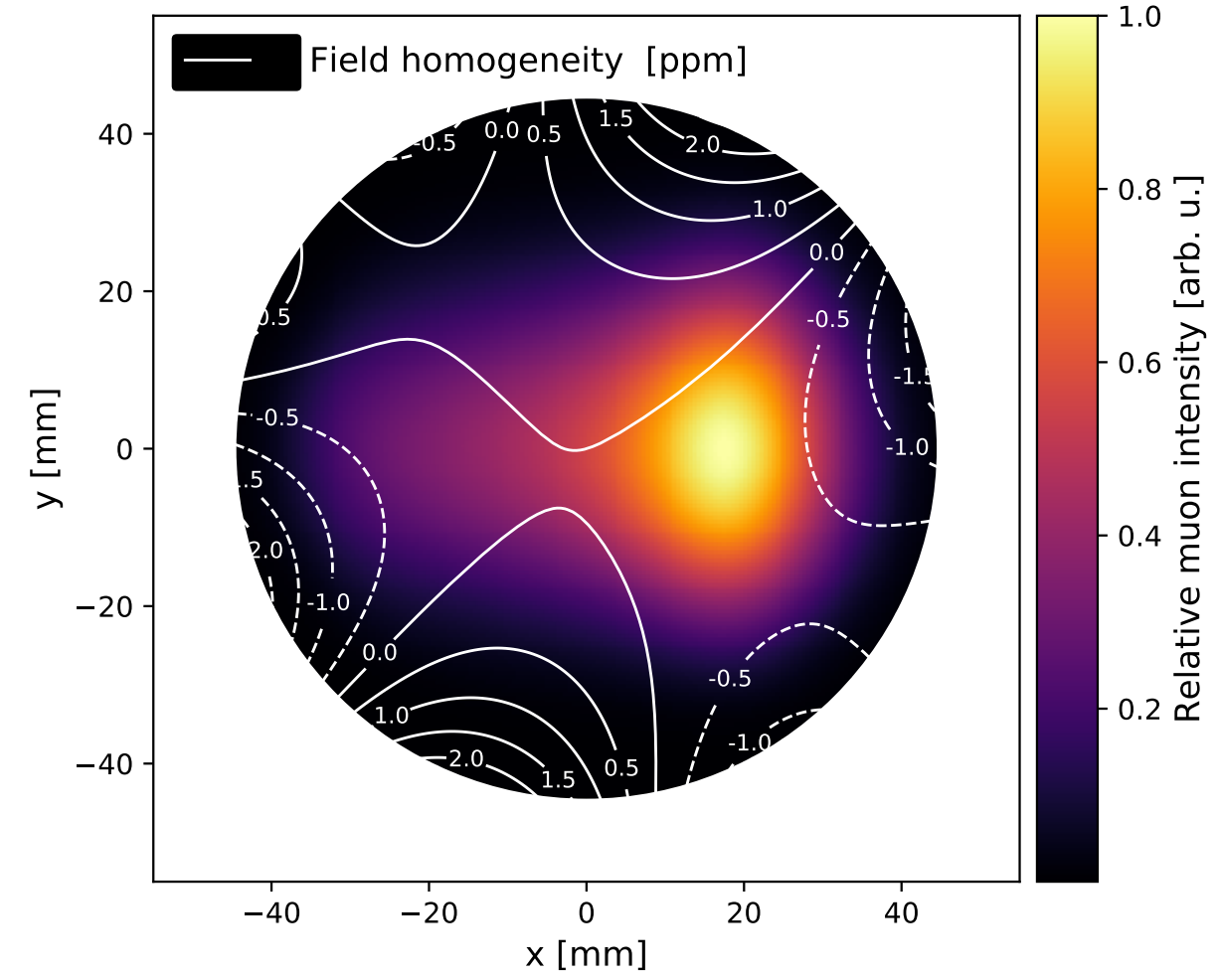}
	\caption{The azimuthally-averaged magnetic field map (level curves) overlayed onto the muon distribution (heat map). The muon distribution is used to weight the magnetic field average to calculate the average field experienced by the muons.}
	\label{fig:field-map-cross-muons}
\end{figure}

The straw trackers are formed of layers of overlapping straws filled with gas that is ionized as positrons travel through the device towards the calorimeters. By measuring the positions where the ionization occurs in the straws, we can find the tracks the positrons took from their decay positions (see Figure \ref{fig:tracker-cartoon}). Extrapolating backwards through the magnetic field, these tracks are used to determine the decay vertices of the positrons, which are used as a proxy for the muon distribution at the two azimuthal locations of the straw trackers. We use the distributions averaged over several hours of data collection as a weighting function when we average the magnetic field's non-uniformity, as shown in Figure \ref{fig:field-map-cross-muons}.

\section{Combining the Measurements}

The uncertainty on the Run-1 result is dominated by statistical uncertainty (\SI{434}{ppb}) compared to the total systematic uncertainty (\SI{157}{ppb}). Table \ref{tab:uncertainty} details the corrections and uncertainties. The statistical precision of this single data set is comparable to that of the entire run of the Brookhaven experiment. The largest systematic uncertainties are the phase acceptance and ESQ field transient. Both of these uncertainties are expected to be reduced in future results: the phase acceptance effect will be reduced by repairs to the quadrupoles, and the ESQ field transient was the subject of a significant systematic study to more precisely characterize the effect over the full storage ring.

\begin{figure}[H]
	\centering
	\begin{tikzpicture}
		\input{results.tikz}
	\end{tikzpicture}
	\caption{The new experimental average of the muon magnetic anomaly is in greater tension with the Standard Model than the previous result from Brookhaven \cite{gm2prl2021}.}
	\label{fig:results}
\end{figure}
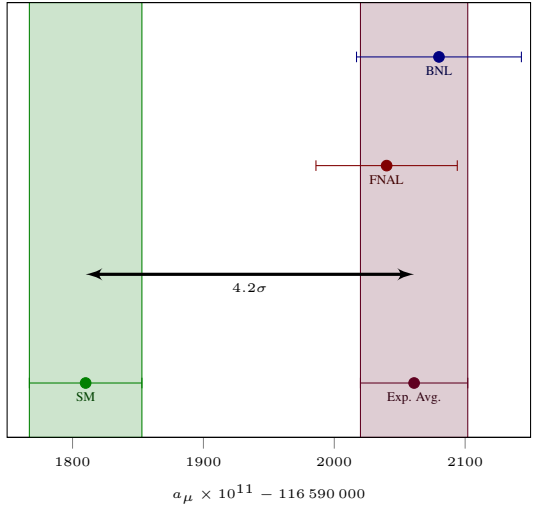

The measurement for Fermilab of $a_\mu(FNAL)=116592040(54)\times10^{-11}$ agrees well with the previous measurement from Brookhaven, as seen in Figure \ref{fig:results}. Because each experiment's uncertainty is dominated by statistics, it is reasonable to combine the two, leading to a new world average of $a_\mu(Exp)=116592061(41)\times10^{-11}$. The reduced error bars due to the higher statistical precision puts the experimental value further in tension with the theoretical prediction, increasing the discrepancy to 4.2 $\sigma$.

Since the completion of Run-1, three more data sets have been collected, Runs-2--4, with Run-5 underway at the time of writing this paper. Run-6 is currently being planned for 2022-2023. These additional runs represent a tenfold improvement in the statistics of the experiment, giving confidence that the statistical uncertainty will be brought down significantly in future results. Additionally, systematic uncertainties are being lowered by a combination of upgrades and studies. Future publications will offer higher precision.

\begin{table}[H]
	\centering
	\begin{tabular}{lrr}
		\midrule\midrule
		Quantity & Correction Terms & Uncertainty \\
		& (ppb) & (ppb) \\
		\midrule
		$\oa$ (stat.) & -- & 434 \\
		$\oa$ (syst.) & -- & 56 \\
		\midrule
		$C_e$ & 489 & 53 \\
		$C_p$ & 180 & 13 \\
		$C_{ml}$ & -11 & 5 \\
		$C_{pa}$ & -158 & 75 \\
		\midrule
		$\oppt$ & -- & 56 \\
		$B_k$ & -27 & 37 \\
		$B_q$ & -17 & 92 \\
		\midrule
		$\mu_p'(\SI{37.4}{\celsius})/\mu_e$ & -- & 10 \\
		$m_\mu/m_e$ & -- & 22 \\
		$g_e/2$ & -- & 0\\
		\midrule
		Totals & 544 & 462\\
		\midrule\midrule
	\end{tabular}
	\caption{The corrections and uncertainties to the terms in Equations \ref{eq:ratios} and \ref{eq:schematical}.}
	\label{tab:uncertainty}
\end{table}

\section{Acknowledgments}

We thank the Fermilab management and staff for their strong support of this experiment, as well as the tremendous support from our university and national laboratory engineers, technicians, and workshops. The Muon $g-2$ Experiment was performed at the Fermi National Accelerator Laboratory, a U.S. Department of Energy, Office of Science, HEP User Facility. Fermilab is managed by Fermi Research Alliance, LLC (FRA), acting under Contract No. DE-AC02-07CH11359. Additional support for the experiment was provided by the Department of Energy offices of HEP and NP (USA), the National Science Foundation (USA), the Istituto Nazionale di Fisica Nucleare (Italy), the Science and Technology Facilities Council (UK), the Royal Society (UK), the European Union's Horizon 2020 research and innovation programme under the Marie Sk\l{}odowska-Curie grant agreements No. 690835, No. 734303, the National Natural Science Foundation of China (Grant No. 11975153, 12075151), MSIP, NRF and IBS-R017-D1 (Republic of Korea), the German Research Foundation (DFG) through the Cluster of Excellence PRISMA+ (EXC 2118/1, Project ID 39083149). 

\end{multicols}
\medline
\begin{multicols}{2}

\end{multicols}
\end{document}

%% file: defs.tex
\newcommand{\beeq}{\begin{equation}}
	\newcommand{\eneq}{\end{equation}}
\let\[\beeq
\let\]\eneq

\newcommand{\amu}{a_\mu}
\newcommand{\oa}{\omega_a}

\newcommand{\opp}{\omega'_p}
\newcommand{\oppt}{\tilde\opp}

%% file: ring-schematic.tikz

\node at (90:0.65) {\scriptsize\ang{0}};
\node at (0:0.65) {\scriptsize\ang{90}};
\node at (-90:0.65) {\scriptsize\ang{180}};
\node at (180:0.65) {\scriptsize\ang{270}};

\foreach \i in {1,...,22,24}{
	\draw (90-15*\i-7.5:1) arc (90-15*\i-7.5:90-15*\i-22.5:1);
}
\foreach \i in {1,...,24}{
	\draw (90-15*\i-7.5:0.95) arc (90-15*\i-7.5:90-15*\i-10.5:0.95);
	\draw (90-15*\i-10.5:0.95) -- (90-15*\i-22.5:0.9);
	\draw (90-15*\i-22.5:0.9) -- (90-15*\i-22.5:0.95);
}
\draw (90-15*23-7.5:1) -- ++(80:0.05) node[] (a) {};
\draw (90-15*24-7.5:1) -- ++(95:0.01) -- (a.center);

\draw[ultra thick, latex'-] (90-15*23-7.5-1+3:1.03) -- ++(170+2.5:0.5) node[anchor=east] {\scriptsize beam};

Inflector
\draw[ultra thick, purple!50!blue] (90-15*23-7.5-1:1.015) -- (90-15*23-7.5-15:.98);
\node[purple!50!blue, anchor=-90] at (90:1.025) {\scriptsize inf.};

Kickers
\foreach \i in {4,5,6}{
	\draw[ultra thick, cyan] (90-15*\i-7.5-3:0.975) arc (90-15*\i-7.5-3:90-15*\i-7.5-15:0.975);
}
\foreach \i in {1,2,3}{
	\node[cyan, anchor=270-15*\i-60] at (90-15*\i-60:1.025) {\scriptsize K\i};
}

Quads
\foreach \i in {1,...,4}{
	\draw[ultra thick, red] (90-90*\i-22.5-3:0.975) arc (90-90*\i-22.5-3:90-90*\i-22.5-15:0.975);
	\draw[ultra thick, red] (90-90*\i-22.5-18:0.975) arc (90-90*\i-22.5-18:90-90*\i-22.5-45:0.975);
	\node[red, anchor=180+90-30-90*\i+90] at (90-30-90*\i+90:1.025) {\scriptsize Q\i S};
	\node[red, anchor=180+90-52.5-90*\i+90] at (90-52.5-90*\i+90:1.025) {\scriptsize Q\i L};
}

Calos
\foreach \i in {1,...,24}{
	\draw[fill, green!80!blue, rotate around={-15*\i-22.5-1.5:(90-15*\i-22.5-0.75:0.94)}] (90-15*\i-22.5-0.75:0.94) rectangle ++(-45:0.05);
	\node[green!80!blue, anchor=90-22.5-15*\i+15-1.5] at (90-22.5-15*\i+15-1.5:0.925) {\tiny\i};
}

Trackers
\foreach \i in {13, 19}{
	\foreach \j in {1,...,8}{
		\draw[thick, blue] (90-15*\i-7.5+\j:0.875+0.003*\j+0.01) -- ++((90-15*\i-7.5:0.06-0.002*\j);
	}
}
\node[blue,anchor=-105] at (-105:0.9) {\scriptsize T1};
\node[blue,anchor=165] at (165:0.9) {\scriptsize T2};

%% file: tracker-cartoon.tikz
\draw [very thick, dotted, color=blue] (-4.9, -0.1) arc (103:74.75:20);
\draw [blue] (-3.66, 0.15) -- ++(0,-1.5) node [anchor=west, blue] {\tiny Muon Orbit};

\draw [very thick, fill] (0,0) circle (0.025);


\draw [ultra thick, cyan] (-13:0.875) -- ++(75:0.375);
\draw [ultra thick, cyan] (-13:0.8375+0.3141) -- ++(75:0.4);
\draw [ultra thick, cyan] (-13:0.8375+2*0.3141) -- ++(75:0.425);
\draw [ultra thick, cyan] (-13:0.8375+3*0.3141) -- ++(75:0.45);
\draw [ultra thick, cyan] (-13:0.8375+4*0.3141) -- ++(75:0.475);
\draw [ultra thick, cyan] (-13:0.8375+5*0.3141) -- ++(75:0.5);
\draw [ultra thick, cyan] (-13:0.8375+6*0.3141) -- ++(75:0.525);
\draw [ultra thick, cyan] (-13:0.8375+7*0.3141) -- ++ (75:0.55);
\draw [cyan] (-13:0.8375+3*0.3141) -- ++(0, -1) node [anchor=west, cyan] {\tiny Straw Trackers};

\draw [thick, cyan, fill=cyan!30!white] (-13:3.42) -- ++(75:0.5) -- ++(-15:0.5) -- ++(255:0.5) -- (-13:3.42);
\draw [cyan] (3.8,-0.88) -- ++(0,-0.5) node [anchor=west, cyan] {\tiny Calorimeter};

\draw [Rays-{Rays[n=7]}, red, thick, dashed] (-3,0.25) arc (92:73:21);
\draw [red] (-2.92,0.25) -- ++(0,1.25) node [anchor=west, red] {\tiny High Momentum Decay Positron};

\draw [Rays-{Rays[n=7]}, red, thick, dashed] (0,0.4) arc (85:60:9);
\draw [red] (0.1,0.4) -- ++(0,0.75) node [anchor=west, red] {\tiny Low Momentum Decay Positron};
	

%% file: results.tikz
\begin{axis} [width=\linewidth, xmin=1750, xmax=2150, ymin=0.5, ymax=4.5, ymajorticks=false, xtick pos=bottom, xtick align=outside, xtick={1800,1900,2000,2100}, xlabel={$\amu\times10^{11} - \num{116590000}$}, /pgf/number format/1000 sep={}, font=\tiny]
	
	\draw [color=green!50!black, fill=green!30!gray!30!white] (axis cs: 1767, 4.5) rectangle (axis cs: 1853, 0.5);
	\addplot+[only marks, mark=*, green!50!black, mark options={fill=green!50!black}][error bars/.cd, x dir=both, x explicit] coordinates {(1810, 1) +- (43,0)};
	\node [anchor=north, text=green!25!black] at (axis cs: 1810, 1) {SM};
	
	\draw [color=purple!50!black, fill=purple!30!gray!30!white] (axis cs: 2020, 4.5) rectangle (axis cs: 2102 ,0.5);
	\addplot+[only marks, mark=*, purple!50!black, mark options={fill=purple!50!black}][error bars/.cd, x dir=both, x explicit] coordinates {(2061, 1) +- (41,0)};
	\node [anchor=north, text=purple!25!black] at (axis cs: 2061, 1) {Exp. Avg.};
	
	\addplot+[only marks, mark=*, blue!50!black, mark options={fill=blue!50!black}][error bars/.cd, x dir=both, x explicit] coordinates {(2080, 4) +- (63,0)};
	\node [anchor=north, text=blue!25!black] at (axis cs: 2080, 4) {BNL};
	
	\addplot+[only marks, mark=*, red!50!black, mark options={fill=red!50!black}][error bars/.cd, x dir=both, x explicit] coordinates {(2040, 3) +- (54,0)};
	\node [anchor=north, text=red!25!black] at (axis cs: 2040, 3) {FNAL};
	
	\draw [latex'-latex', very thick, black] (axis cs: 1810,2) -- (axis cs: 2061,2);
	\node [anchor=north, text=black] at (axis cs: 1810/2 + 2061/2, 2) {$4.2 \sigma$};
	
	\draw (axis cs: 1750, 4.5) -- (axis cs: 2150, 4.5);
	\draw (axis cs: 1750, 0.5) -- (axis cs: 2150, 0.5);
	
\end{axis}

%% file: RADpyC-conf-proc.bbl
\begin{thebibliography}{99}
	
\newcommand{\etal}{\emph{et. al.}}

\bibitem{gm2prl2021} B. Abi \etal, Phys. Rev. Lett. \textbf{126}, 14 (2021).
\bibitem{gm2omegaa2021} T. Albahri \etal, Phys. Rev. D \textbf{103}, 7 (2021).
\bibitem{gm2field2021} T. Albahri \etal, Phys. Rev. A \textbf{103}, 4 (2021).
\bibitem{gm2bd2021} T. Albahri \etal, Phys. Rev. Accel. Beams \textbf{24}, 4 (2021).
\bibitem{Bennett2006} G. Bennett \etal, Phys. Rev. D \textbf{73}, 7 (2006).
\bibitem{Jackson} J. D. Jackson. Classical Electrodynamics, Third Edition. Wiley (1998).
\bibitem{gm2whitepaper2020} T. Aoyama \etal, Phys. Reports, \textbf{887}, (2020).
\bibitem{Roberts2018} B. Roberts, SciPost Phys. Proc., Proc. 15th Intl Workshop on Tau Lepton Physics (2018).
\bibitem{Farooq2020} M. Farooq \etal, Phys. Rev. Lett. \textbf{124}, 22, (2020).

\end{thebibliography}
